# Why do developers take breaks from contributing to OSS projects? *A preliminary analysis*

Giuseppe Iaffaldano
*Department of Computer Science*
*University of Bari*
Bari, Italy
giuseppe.iaffaldano@uniba.it

Igor Steinmacher
*School of Informatics, Computing, and Cyber Systems*
*Northern Arizona University*
igor.steinmacher@nau.edu

Fabio Calefato
*Ionian Department*
*University of Bari*
Taranto, Italy
fabio.calefato@uniba.it

Marco Gerosa
*School of Informatics, Computing, and Cyber Systems*
*Northern Arizona University*
marco.gerosa@nau.edu

Filippo Lanubile
*Department of Computer Science*
*University of Bari*
Bari, Italy
filippo.lanubile@uniba.it

***Abstract*—Creating a successful and sustainable Open Source Software (OSS) project often depends on the strength and the health of the community behind it. Current literature explains the contributors' lifecycle, starting with the motivations that drive people to contribute and barriers to joining OSS projects, covering developers' evolution until they become core members. However, the stages when developers leave the projects are still weakly explored and are not well-defined in existing developers' lifecycle models. In this position paper, we enrich the knowledge about the leaving stage by identifying sleeping and dead states, representing temporary and permanent brakes that developers take from contributing. We conducted a preliminary set of semi-structured interviews with active developers. We analyzed the answers by focusing on defining and understanding the reasons for the transitions to/from sleeping and dead states. This paper raises new questions that may guide further discussions and research, which may ultimately benefit OSS communities.**

## I. Introduction

The open source software phenomenon raises an innovative process, i.e., invention, innovation and diffusion [1], which allowed volunteer developers to build high-quality software supported by online communities [2]. Thus, building a successful open source software project and keeping it up to date depends on the strength and the health of the community behind it [3], [4]. Since developers play a key role, it is important to understand their life cycle and make sure that a workforce and knowledge base remains in the community.

So far, the OSS-related literature frames the developers' lifecycle focusing on how people join the projects [5], including the barriers they face [6], and how they grow and become long-term contributors and core members [7]–[9]. Although seamlessly important, there is no well-defined lifecycle model encompassing the stages when developers leave the projects. Most of the effort about this phase has been put on understanding the risks in which projects incur when losing developers [10]–[12], and on understanding the factors related to the developers' abandonment through the survival analysis technique [13]. Avoiding developers to leave the project is a matter that affects communities and has interested researchers who investigated strategies to keep developers engaged [14]–[18]. However, general retention strategies are often not enough to make the projects grow in the right way. On the one hand, project turnover helps to keep OSS projects alive and brings fresh energy as well as new ideas in the community [19]. On the other hand, it may disrupt the community and lower the product quality [15], [20].

By lurking in some projects on GitHub, we noticed that some developers take long breaks from development, while others suddenly disappear from the contribution timeline. We came up with metaphors suggesting that developers may spend some time sleeping or they can die. So, in this position paper, we explore the phenomenon of developers becoming inactive or abandoning the projects. To do so, we introduce the concepts of *sleeping* and *dead* developers, representing those developers who take temporary or permanent breaks from contributing code to the projects.

With this position paper, we want to open a discussion around this topic and bring evidence of the reasons why developers leave the projects and of the signals to help to identify that this phenomenon is happening.

In particular, we present the analysis of a set of semi-structured interviews with developers with core roles in different OSS projects. Based on that, we define *sleeping* and *dead* statuses as metaphors for developers who stay away from the project for a while, and for those who abandon the project. We also identify a set of motivations that trigger the transition to these states, which varies between personal and project-related reasons. Besides, we come up with some questions and hypotheses to enable an in-depth discussion about the topic.

The rest of this paper is structured as follows: in the next section (II), we point out the main studies in the field and the gap we want to fill up; in Section III, we explain the method used to conduct this preliminary study. The results of the interviews analysis are reported in section IV. Finally, in Section V, we discuss the main findings and raise questions to be further explored in future work. Conclusions are reported in Section VI.

## II. Related Work

For the past years, researchers have been investigating the role and evolution of developers in OSS communities. Part of this research focuses on the joining process, investigating the steps for developers to become active members or to reach the core of a project. For example, Nakakoji et al. [21] proposed the *onion model*, aiming to represent the general structure of OSS and the process that developers follow to become core members. Von Krogh et al. [5] proposed a joining script, based on steps that developers need to follow to become part of the project. Ducheneaut [8] analyzed mailing list archives, offering an in-depth look at a successful newcomer's socialization history. From a different perspective, Steinmacher et al. [22] explained the joining process using two stages, namely *onboarding* and *contributing*, and identified the forces that push developers towards a project (motivation [23], [24] and attractiveness [19], [25]) and those that hinder developers' onboarding [26].

While all the aforementioned studies discuss how to become a successful member of a community, some others focus on understanding the potential issues of developers leaving OSS projects. Some of these studies discuss and operationalize the so-called *truck factor* [10]–[12], which is defined as the number of people that have to be hit by a truck (i.e., quit) before the project itself is at risk [27]. For example, Avelino et al. [11] computed the truck factor for popular GitHub projects and found that nearly two-thirds of them depend on one or two developers to survive.

Prior research has also investigated motivations for developers to leave or stay in the projects. For example, Lin et al. [13] studied why some developers are more likely to continue their contributions than others. Several studies try to understand the motivation behind project failures using survival analysis [28]–[32]; however, only a few studies shed light on the antecedents of users' leaving that caused the failure of OSS projects (e.g., [33], [34]). Furthermore, to the best of our knowledge, no previous study has investigated whether and for how long developers who supposedly abandoned projects resume contributing.

In summary, existing research focused on studying the health of OSS communities concerning the developers' lifecycle, including onboarding, retention, and turnover. In this paper, we set out to broaden such lifecycle by modeling an intermediate state where developers pause their active contribution and then resume it, or rather extend their hiatus until eventually abandoning a project altogether.

## III. METHOD

Modern platforms like GitHub[1] or GitLab provide graphs that display the timeline of the contributions made by a developer to a project. By observing these timelines of contribution, one can notice a particular frequency of commits, or *rhythm* of a developer (Fig. 1a). We manually inspected several graphs and realized that sometimes these rhythms change, with developers drastically reducing the frequency of their contributions or even disappearing for a while, i.e., developers take breaks from contributing. In the example shown in Fig. 1b, we can observe the contribution timeline of a developer of the *rails* project: contributions peaks between 2004-2009 and are drastically reduced until they break at the end of 2009; then, the developer returns after about six months.

To model these periods, we took inspiration from the *sleep stages*, comparing the commit timeline to the circadian rhythm (or sleep-wake cycle). The *wake stage* is when people are alert and wakeful with intense brain activity and corresponds to the state which developers are actively contributing code to a project repository. During *sleep stage,* people's brain activity is largely reduced, but other life signals are still present. In our metaphor, this stage corresponds to the state in which developers pause their commit activity, but they still provide other signals of presence in the community, such as commenting issues, reviewing pull requests, and sending emails to mailing lists. Like individuals normally transition from *sleeping* back into the *wake stage*, developers also typically resume their commit activity in OSS projects after some time. However, with the indefinite extension of the *sleep stage*, life signs may progressively decline and eventually culminate with the *dead stage*. Similarly, if developers' hiatus continues, with their activity further reduced down to the complete lack thereof, they can be presumed *dead* in the sense of having abandoned the project.

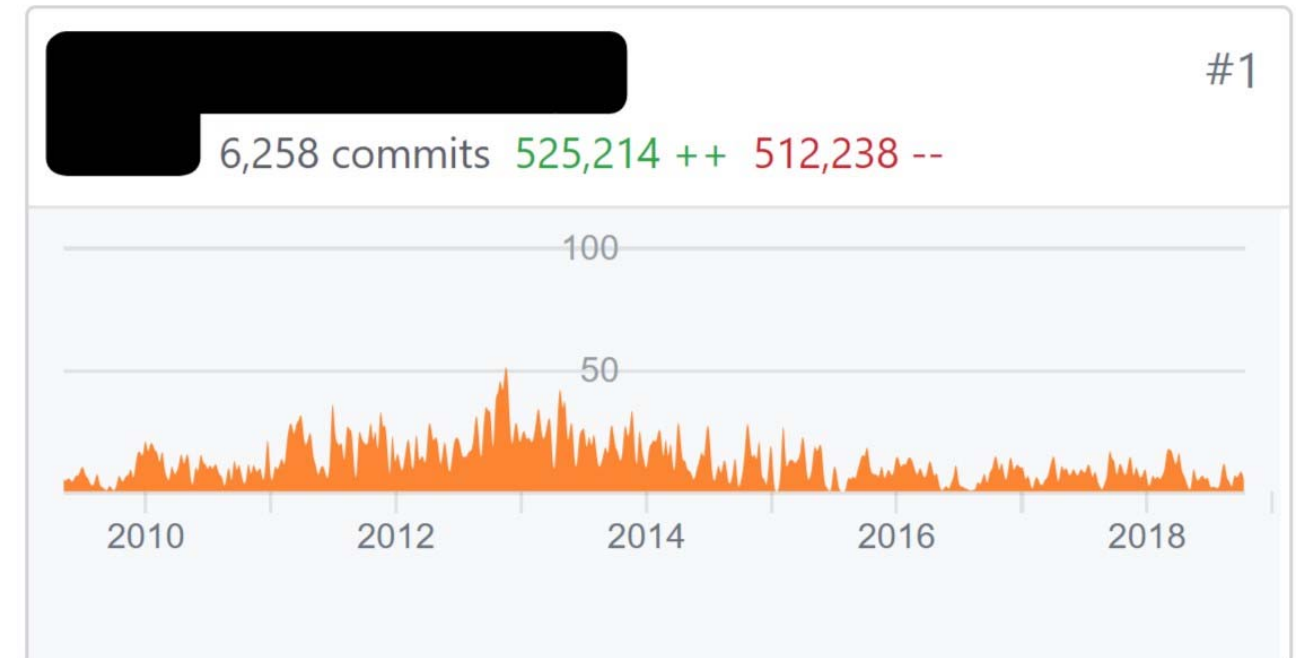

Fig. 1a. Timeline of a frequent contributor

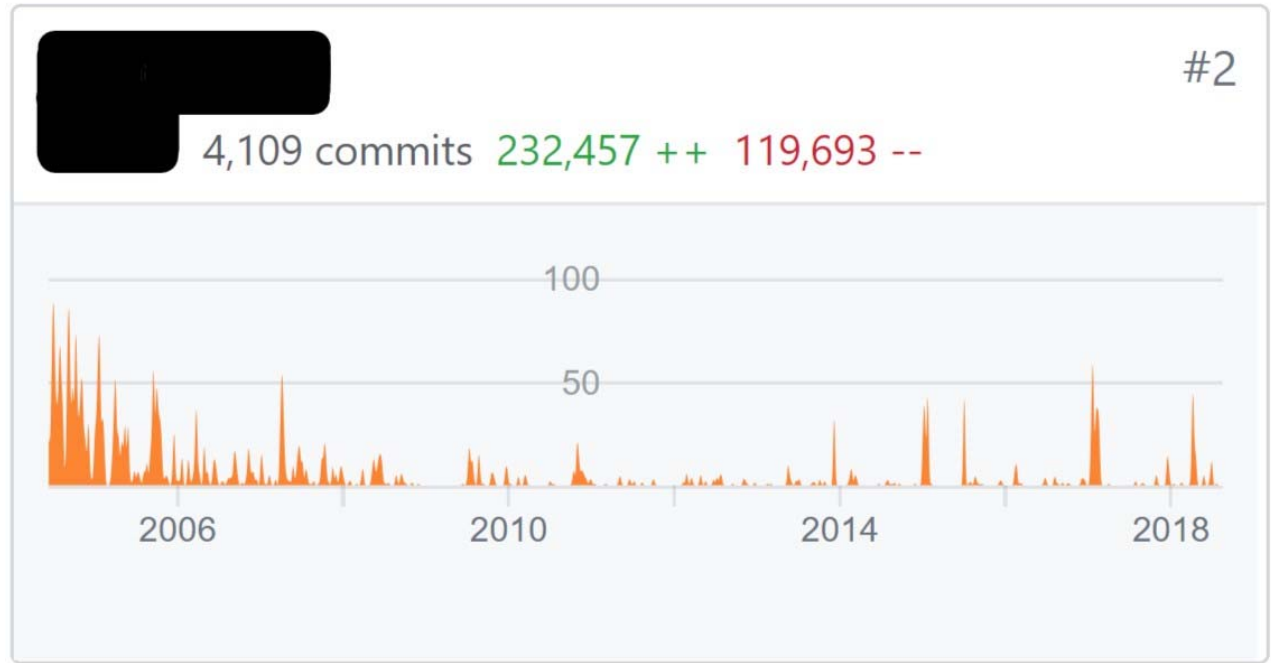

Fig. 1b Timeline of a contributor interrupted for few months

To further our understanding of why developers may take breaks from contributing to OSS projects and for how long before being considered *dead*, we collected the point of view of OSS developers. Accordingly, we formulated the following research questions to guide our study:

*RQ1: How do OSS developers define sleeping and dead states in OSS projects?*

*RQ2: What are the factors that drive developers to sleep or die?*

*RQ3: What are the signals indicating that developers are going to sleep or die?*

To answer these questions, we conducted a qualitative study, using semi-structured interviews. We interviewed 6

[1] Contributors' timeline for the Linux project: https://github.com/torvalds/linux/graphs/contributors

developers (5 male and 1 female) with a history of contributions to several OSS projects; 5 of them have been maintainers or core members, and 1 of them works in a company that owns different OSS projects. The interviews were conducted using text-based chat and lasted for 60 to 90 minutes.

Each interview was structured in two main parts, plus a third conclusive one. In the first part, we collected information about the developer's experience with OSS projects and contribution frequency; in the second part, we collected the developer's perception about '*sleeping*' and '*dead*' developer concepts, the reasons to move to and from these states, examples, and hints about how to identify them. In the third part, we asked about the communication and coordination means used in the projects, and also about the perceived reputation change of a developer who returns contributing after a *sleeping* or *dead* period.

To analyze the answers, we applied an open-coding strategy. First, one of the authors added the transcripts into spreadsheets and started to identify common themes related to the developer lifecycle in OSS projects supported by a more experienced researcher. Then, we refined the classification of the concepts in more specific categories and performed card sorting. Finally, we discussed in iterative consensus meetings to clarify ambiguities and consolidate the categories.

## IV. RESULTS

As a result of the interview analysis, we derived a 'cyclic' model of developers' states that also includes contribution breaks (Fig. 2). The model is a state diagram in which circles represent states, namely *active*, *sleeping*, and *dead*, and arrows represent transitions between them. Intuitively, we define any developer who is currently committing to a project to be in the *active* state. In the rest of this section, we discuss and provide the definitions for the *sleeping* and *dead* developers, as they emerged from the interviews; we also explain the transitions between the states by exploring the main reasons that drive developers to change their state, as reported by the interviewees.

### *A. Definitions of Sleeping and Dead states.*

The interviewed developers agreed with the idea of *sleeping* developers being those who do not contribute code but still show interest in the project in other ways such as answering emails and participating in discussions. For example, developer D2 mentioned a project where he *"[...] participated very little, just following some discussions and being present in some meetings. So, I was sleeping." Sleeping* developers keep following the project's evolution because they are waiting for the right moment to contribute again: *"Sometimes a person does not interact because [the person] doesn't feel that is right to intervene in something that they cannot actually act [upon] at the moment, but they continue to follow and care about the project"* (D2). When active developers are *sleeping* the rest of the community expect them to get back contributing soon, as mentioned by D4: *"if that person is an active contributor for a very long time, then the expectation is high from that person. If a developer has been very active, as a manager, you tend to see the glass half-filled and expect a new contribution from that person. So, as a project manager, I would consider that person sleepy."*

Similarly, the analysis of the interviews helped us to define *dead* developers like those who not only have stopped providing code contributions for some time (D3: "*after a period of no activity, you can consider one dead"*; D2: "*[they] will not contribute anymore to that community*"), but also do not participate in any other community activity (D2: *"[dead is when I] don't wanna come back to interact with the project"*). One developer even provided insight about the length of such contribution break before one project member can be considered *dead*: *"absolute inactivity in the last year (considering all the activities of the project)"* (D3).

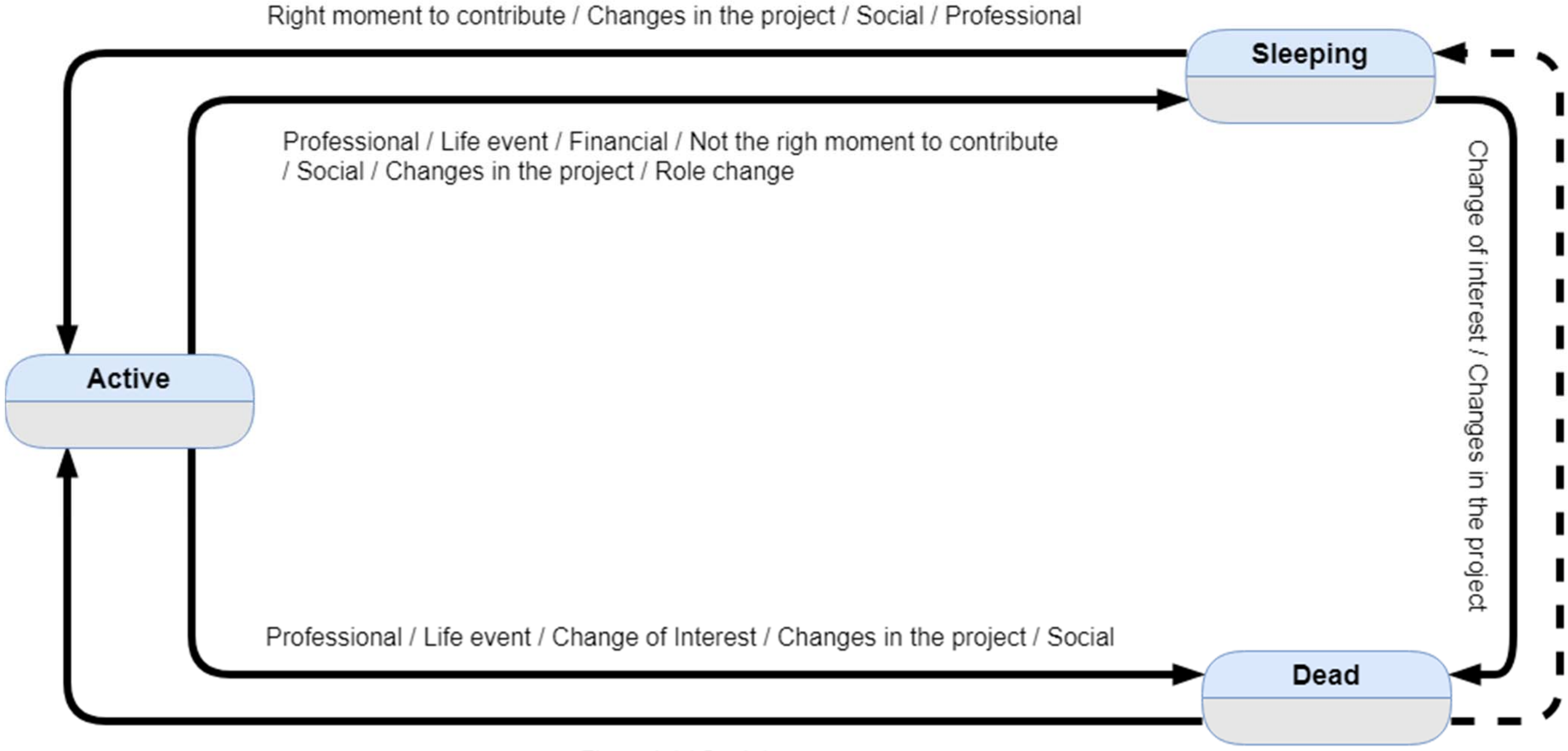

Fig. 2. State diagram showing the *active*, *sleeping* and *dead* states, and the transitions between them. Solid lines are used for transitions emerged during qualitative analysis, the dashed line represents the hypothesized transition from *dead* to *sleeping* state.

In addition to understanding the meaning of *sleeping*, we also investigated the reasons (or triggers) that make the developers transition from one state to another, which are reported in the following subsections.

### B. Reasons for going to sleep

The interviewees mentioned diverse reasons as for why developers go in the *sleeping* status: *"I can say that I am sleeping if I have some external problem that takes my availability (as other work) or humor to contribute (personal problem) or when I had some trouble with a community member, but as soon as possible I want to come back."* (D2). We analyzed all the reasons provided during our interviews, and further categorized them into personal and project (or community-related) reasons. We synthesize all the reasons that emerged from the interviews in TABLE I.

#### 1) Personal reasons

Since several OSS developers are volunteers or work part-time on OSS, we understand that developers' attention may be taken from other professional priorities: *"It is more common [to go sleeping] when the developer is not paid to contribute to that software when it is developed in free time"* (D5). For example, a student who contributes to an OSS project may be overwhelmed by academic activities, especially during the exams period. One example is something that happened with D4: *"I am currently inactive due to my exams […] I mailed to the community that I have not been active for that, and once they know my exams are over, they will expect some commit."* Even someone who works in a different context may be forced to focus on something different than the project: *"I think the sleeping is when the developers are busy for something with more priority level"* (D5).

Usually, a personal life event is enough to take the developers' attention. As D5 mentioned: *"It is common to see developers saying in mailing lists 'hello people, I will be off for some months because I am a father now/I am on vacations/I am sick'."* Sometimes, volunteer developers need to dedicate more time to the work for financial problems: *"I have friends that are sleeping for financial issues"* (D2). It often happens that students contribute in some projects for the Google Summer of Code (GSoC), but they go *sleeping* for the school period and come back the following summer. One of the interviewees who takes care of the contributions of his students in OSS said: *"I have a lot of experience with sleeping developers because I participated in the GSoC and typically you have some students who are very active in participating during the summer, and then during the school they are too busy with the school work and they don't really have time to continue with the project […] sometimes they come back the next summer to work on the same project […]. I think it is quite frequent to have these sleeping developers that are just coming back in the summer because the GSoC only pays to work on summer"* (D6).

#### 2) Project-related reasons

On the other side, some factors are directly related to the project and bring developers to take a break from contributing. Often, it is not the right moment for developers to give their contribution as mentioned by D2: *"Sometimes a person does not interact because [the person] doesn't feel that is right to intervene in something that they cannot actually act at the moment, but they continue to follow and care about the project in silence."* For example, there may be other stakeholders taking care of the project: *"We got some funding, and my colleague decided for a strategy of contracting 'internships' He was responsible for managing the internship of the moment. So, I participated very little, just following some discussions and being present in some meetings. I was sleeping"* (D1).

Sometimes, the social behavior of the community behind the project makes the difference. Being reactive and giving feedback may help new developers to feel welcomed in the community and give more frequent contributions [35], [36]. Ignoring contributions may drive developers away, as happened to D3: *"I now consider myself a sleeping developer […] After sending at least a 60 of commits in various PRs and not yet having received a reply, I preferred temporarily not to pour out more extra time in the project."* It is also important to develop a sense of community that helps people working better together, avoiding personal problems. D2, for example, although interested in the project, mentioned that *"I'm sleeping in [project] […] I can say that I have some disagreements with the community. But I like to develop and improve the software. I have friends that are sleeping for the same reason*."

TABLE I. REASONS TO STOP CONTRIBUTING TO A PROJECT

| Category | Main Reason | Examples |
|---|---|---|
| Personal | Professional | Courses for students; exams |
| | Life event | Death; Child birth; Sickness |
| | Financial | GSoC; need time to work |
| | Change of interest | Lack of interest; Migration |
| Project | Not the right moment to contribute | Does not feel responsible; Delivered feature; Others taking over |
| | Social | Problems between members; not receiving feedback |
| | Changes in the project | Technical; organization (governance) |
| | Role change | Becoming a project manager |

Moreover, even if necessary for the evolution of the project, big changes in the project may find developers not prepared to face them and drive them to take a break to adapt and reconsider their work strategies: *"I'm sleeping in [project] because I like the project and the community and wanna come back, but I think I need more time to understand the current architecture, build my development environment and probably I will need to improve my technical skills."* (D2).

Finally, a simple change of role of a developer may be lead to a break from contributing code. When developers get the role of maintainer, they may initially be more focused on very different tasks than giving code contributions: *"In the beginning I contributed fixing small bugs, doing small changes in the code. After some time, I made big contributions. After that, I started to work more in the management side and mentoring newcomers. Since then I am working more in the management side. When I talk about the management side, I am saying triaging bugs, writing new issues, feature requests, assigning developers to tasks, reviewing changes, and more"* (D5).

### *C. Reasons for awakening*

The *awakening*, i.e., the return to activity after a *sleeping* period, usually happens when the motivation for which the developers have stopped their contributions is no longer valid, and they return to commit. A motivation to return may be due to a need, or interest in the project (e.g., fixing a bug, developing a new feature). D3 reported mentioned: *"a person was interested in integrating a feature in [project] initially it was quite active, then for 4-5 months it disappeared and then came back [...] the return happened because, from what I understood, his company is interested in using [the project]."*

As we mentioned before, once the *sleeping* developers perceive that their knowledge is required to work an issue, they wake up: *"my contribution was expected at some point. And we had a problem too hard for the internships to solve... none of them were able to resolve it, so I woke up to solve this problem. I solved. And I came back to sleep"* (D1). If *sleeping* developers do not realize that the community needs their support, maintainers may ask them to help: *"I think if someone wakes me up, I will return contributing"* (D2).

When developers are *sleeping* because they disagree with changes in the project (governance strategies) or they have some social concerns (conflicts with community members), they would be driven to contribute again by some changes in the community or in the project governance. This was mentioned by D2: *"If I see some changes in the community decisions, or if I need to work indirectly, I would return contributing*."

### *D. Reasons for dying*

Dying is the transition to the *dead* state with no more contributions to a project. The developers provided during their interviews similar reasons as those reported for going to *sleep*. In this case, most of the reasons are personal.

One interesting finding is that, sometimes, the *sleeping status* is a temporary condition before developers' death. If developers do not contribute for a long time, they may slowly lose interest in the project and slow down their activities until they lose the awareness and the knowledge on the project (D3: *"if the project is very active, the knowledge of the developer quickly disappears"*).

#### *1) Personal reasons*

Professional-related factors may also drive developers to leave the project: *"A colleague who left the company was very active in a couple of side-projects (not exactly related to the company), but it's been quite a while since he's not working there, I suppose it's due to the change of job and the consequent lack of time" (D3*). There are some cases when students are contributing and leave the projects because they had been asked to contribute as part of a course. When the course comes to an end, they do not continue contributing – *"we do not expect students to help after finishing the course [...]" (D1)* – or after they graduate and get a job *"they graduate, they get a job, and they are not using to stay in the software anymore, so they don't contribute anymore. That's another dead developer" (D1*).

Among the personal reasons, we also found critical life events that usually drastically change the lifestyle of developers. D1, for example, mentioned one developer who *"[...] helped me in our software. It was a lot of work. Now, I do not expect he help (unless I ask about specific topics), he got a child now*." D5 mentioned sad events *"when something critical happens [...], like death itself*."

The main and most generic reason that an interviewee gave us as for leaving is the loss of any interest in the project: *"The dead mode happens when the developer finds something more interesting to work and leaves the development of previous software"* (D5).

#### *2) Project-related reasons*

Some developers are only interested in a feature or a specific issue, and they will not contribute again: *"it often happens that someone opens an issue and is only interested in that issue"* (D3). Sometimes, arguments with the governance model may drive a developer to decide to leave permanently, as reported by D3: *"another case of dead developer is me in [project]. For lack of time and the management of the project [...] I no longer contributed"*). Finally, one common reason to leave a community is the lack of ties and commitment due to the lack of communication (D1: *"In the case of [project] we have 5 main contributors along the history. One of them is very shy and do not communicate too much. He helped a little for some time and stopped contributing"*).

### *E. Reasons for resurrection*

The *resurrection* from the death seems to be a good metaphor because it is an unexpected event. The interviewed developers mostly did not find good reasons for developers to return to the project once they are *dead*, neither they found an example in their experience: *"In my experience, I don't think there are resurrecting developers. If they decide to move on, they don't really come back"* (D6). If a *resurrection* happens, it is due to a strong motivation, i.e., money or being pushed by something stronger than the decision to leave. D5, for example, mentioned that "*[someone would return] from the dead mode, maybe just if the developer is hired to work on that."*

Sometimes, strong social ties with project members influence the developer choice and trigger a *resurrection*. D1 mentioned a personal case: *"I got an issue that I thought he* *could help so I asked him, and he helped. So, he is the kind of dead contributor that can be resurrected if summoned by me. [...] He helped again because I detected an opportunity where his skills could suit, and I asked for help. It's the kind of appeal: 'I know you have more skills than others in this subject, so we need your help.'"*

### F. Signals

During our analysis, we could also identify signals (or signs) to determine in which state a developer is. These signals came up when we followed up the questions about sleeping and *dead* states, asking the interviewees to give us hints that may help to identify developers going into sleeping or *dead* states. The goal of investigating these signals is to have proxies to identify if developers are going to *sleep*, *die*, *wake* up, or *resurrect*.

We observed that most of the signals are external to the working platform, as they are received via email and private messages: *"their [developers'] interest for the project is a signal. Whenever there is an exam, I mailed to the community that I am not been active for that"* (D4). However, we noticed few signals that may help understanding if the community is going to lose a developer. For example, some interviewees stated that signals indicating that developers are *going to sleep* could be a drastic reduction of the commit rhythm, as stated by D3: *"[Their] frequency of development is decreasing and turns inconstant;" "sleeping: drastically reduced activity (<20% of the usual activity) in the last 2 months."*

It is more challenging to define signals indicating that developers are *dying*, especially because this may happen suddenly and without any warning. However, there are cases of external events mentioned by the interviewees, which anticipate that someone is about to leave the community. For example, when a course that requires a contribution to OSS projects ends, one can anticipate that many students are going to leave the project, probably forever, as mentioned by D1: *"we do not expect students to help after finishing the course."*

Lack of communication is another signal that the project is losing that developer: *"When the frequency of posts decreased and after some years there is no news coming from that developer, it is a strong signal he is now in dead mode"* (D5). In other cases, the developers explicitly warn the community about their intention to leave: *"In mailing lists, it is common to see developers talking publicly about his departure of some project..."* (D5). In this case, it would be easier for the maintainers to plan and, if necessary, replace that developer.

We also identified some signals to help understanding if a developer may be *waking up* or even *resurrecting*. Usually, sleeping developers make the community aware of their return by communicating that they will take care of an issue in the issue tracker. Regarding *dead* developers, since they have been away for a long period and did not participate in any community activity, they give signals of their return by reconnecting to the community via communication channels, for example, *"a comment in an issue, a post in a forum… I would say activities that do not take long"* (D3). D5 reinforced that *"maybe when someone starts to talk again about the project, maybe in some e-mail, or some review..." Dead* developers also need to understand if changes to the project governance happened and adapt their behavior to the new rules. In other words, signals indicate that a *dead* developer becomes a *sleeping* developer before definitively *waking up*.

## V. DISCUSSION

Based on the analysis of the interviews, it was possible to preliminarily define the *sleeping* and *dead* states and the reasons why developers move to and from them. Our results also bring insights and potential discussion on research avenues that need to be further explored.

**How common is this phenomenon?** Given that our interviewees acknowledged the existence of the states we are investigating, it becomes interesting to further the understanding of the transitions to these states to avoid deaths and maintain OSS communities in good health. Thus, exploratory questions aiming to characterize these phenomena become important, such as:

- How common are the transitions to *sleeping* and *dead* states?
- How common are *awakenings* and *resurrections*?

**How to identify sleeping and dead developers.** To answer the previous questions, it is necessary to objectively define dead and sleeping states concerning their inactivity period. The first challenge is to decide between fixed and variable thresholds. Although fixed thresholds make the identification easier, we believe that a variable approach should be more appropriate. We advocate that developers have different contribution rhythms, and therefore their absence may be felt after different periods. For example, while an active developer who contributes every day would be considered *sleeping* after one week and *dead* in two months, another who contributes every ten days would be considered *sleeping* after the same two months. The challenge is, then:

How to have a fair way to identify people in these states, considering different developers' behaviors?

Also, to distinguish *sleeping* and *dead* states, one key element identified was that *sleeping* developers still show interest in the project, (i.e., show some 'vital signs'). As stated by D2, *"[...] this status is a union of factors."* Therefore, it is necessary to investigate these signals of activity in an OSS project environment. Accordingly, we asked interviewees which communication and coordination tools they use, other than mailing lists and issue trackers, and they mentioned blogs, wikis, IRC channels, and even Slack and Telegram (D3: *"Slack and Phabricator";* D4: *"IRC channel for monthly meetings";* D5: *"Blog posts on Planet [...] Some communities use IRC too, or other chat platform like Telegram."*). With these tools, some signals are visible to the community (e.g., answering questions, reviewing code, and following issues) but others are not (e.g., reading issues and mailing list messages, and communicating privately with some members). Thus, another challenge related to the identification of *sleeping* developers is uncovering the potential ways to find their 'vital signs,' i.e.:

Are there ways to identify both visible and invisible signals by analyzing data from issue trackers, mailing lists, and forums?

**Predicting death.** Also related to signals, as presented in Section IV.F, we could identify some of them (e.g., lack of communication or explicit warnings for the community) related to the transitions to the *dead* state. Thus, by anticipating these transitions, it may be possible to prevent developers from *dying*. However, it is not clear whether the signals are good and enough indicators, as reported by D2: *"I think is hard to say that someone is dead because [they] didn't interact to the project, but I agree that the time size can be a good factor."* Thus, it is important to understand

> Which signals would be good predictors that developers are going to *sleep* or *die*?

**Death and turnover: good or bad?** Although researchers see turnover as a threat to the sustainability of projects [15], [20], one of our interviewees said that it might instead bring freshness: *"It is good to keep [old] developers, but the new ones bring energy and new things to the project. [On the other hand] old developers, in general, are very resistant to changes, while new developers can break all the software just to try new things."* (D5) As such, it would be interesting to understand when turnover/death is a good thing.

Moreover, we found that some maintainers are more likely to leverage turnover rather than try to call back developers who left, whereas others are more likely to try *resurrecting* developers. Therefore, a couple of questions are raised:

> - What is the best balance between turnover and retention?
> - In what cases one is preferable to the other?

Since one of the reasons to go to *sleep* is that developers think it is not the right moment to contribute because someone else is taking over, the number of developers that can take care of issues in the project may be relevant – *"If someone is inactive, then anyone who has knowledge do their work. Because its open source anyone can contribute" (D4).* Hence, another interesting question is:

> How does the size of the project influence the likelihood of the transitions to the *sleeping* and *dead* states?

**Is it hard to resurrect?** During the absence of some developers, projects continues to grow and may undergo several changes. For example, the active and core members may have changed, the architecture may have undergone substantial refactoring, or even the organizational/governance structure may be different. All these facts can make it harder for someone who was outside for an extended period to onboard again. Thus, an interesting research avenue is answering the following question:

> What challenges do *dead* developers face when they want to return *active* in a project?

A related aspect would be to better understand the changes in developers' reputation during the dead period and the effect of their reputation on the *resurrection* process.

**Sleeping as a transition to death.** From our results, it is possible to observe that the reasons to *die* or to *sleep* have a large intersection (e.g., professional-related, life events, social-related). From these preliminary interviews, given a reason, it was not possible to understand what would make the transition be to *dead* or *sleeping* state. Besides, this same observation can make one think about *sleeping* as a transitional state to the *death* and makes us raise the following question:

> Is the *death* of a developer preceded by a *sleeping* period?

This is an interesting topic to investigate to make communities aware of potential deaths, so they may come up with ways to awake people who may *die*.

## VI. Conclusion

In this paper, we explored the phenomenon of developers becoming inactive or abandoning the projects. We believe that this topic is important and deserves further investigation since the developers' transition to the *sleeping* and *dead* states may affect the health of OSS communities. By defining these states and identifying reasons that lead developers to move to them, we aim to open the discussion about the topic. The preliminary evidence and questions suggest the necessity of further research, which may ultimately benefit OSS communities.

## Acknowledgment

We are grateful to the software developers who attended the interview, including Filipe Saraiva, Toby Dylan Hocking, Gourav Sardana, Valerio Cosentino, and others who preferred to remain anonymous.

This work is partially funded by project "Creative Cultural Collaboration" (C3), under the Apulian INNONETWORK programme, Italy, CNPq (Grant #430642/2016-4), and FAPESP (Grant #2015/24527-3).

## References

[1] J. A. Schumpeter and R. Opie, "The theory of economic development; an inquiry into profits, capital, credit, interest, and the business cycle,". Harvard University Press, 1934.

[2] B. Fitzgerald, "The Transformation of Open Source Software," *MIS Q.*, vol. 30, no. 3, pp. 587–598, 2006.

[3] K. Crowston and J. Howison, "Assessing the health of open source communities," *Computer (Long. Beach. Calif).*, vol. 39, no. 5, pp. 89–91, May 2006.

[4] G. J. P. Link and M. Germonprez, "Assessing Open Source Project Health Twenty-fourth Americas Conference on Information Systems," 2018.

[5] G. von Krogh, S. Spaeth, and K. R. Lakhani, "Community, joining, and specialization in open source software innovation: a case study," *Res. Policy*, vol. 32, no. 7, pp. 1217–1241, Jul. 2003.

[6] I. Steinmacher, M. A. Graciotto Silva, M. A. Gerosa, and D. F. Redmiles, "A systematic literature review on the barriers faced by newcomers to open source software projects," *Inf. Softw. Technol.*, vol. 59, no. C, pp. 67–85, Mar. 2015.

[7] Q. Hong, S. Kim, S. C. Cheung, and C. Bird, "Understanding a developer social network and its evolution," in *2011 27th IEEE International Conference on Software Maintenance (ICSM)*, 2011, pp. 323–332.

[8] N. Ducheneaut, "Socialization in an Open Source Software Community: A Socio-Technical Analysis," *Comput. Support. Coop. Work*, vol. 14, no. 4, pp. 323–368, Aug. 2005.

[9] M. Zhou and A. Mockus, "What Make Long Term Contributors: Willingness and Opportunity in OSS Community," in *Proceedings of the 34th International Conference on Software Engineering*, 2012, pp. 518–528.

[10] F. Ricca and A. Marchetto, "Are Heroes common in FLOSS projects?," in *Proceedings of the 2010 ACM-IEEE International Symposium on Empirical Software Engineering and Measurement - ESEM '10*, 2010, p. 1.

[11] G. Avelino, L. Passos, A. Hora, and M. T. Valente, “A novel approach for estimating Truck Factors,” *IEEE Int. Conf. Progr. Compr.*, vol. 2016–July, no. Dcc, pp. 1–10, 2016.

[12] M. Ferreira, M. T. Valente, and K. Ferreira, “A Comparison of Three Algorithms for Computing Truck Factors,” in *2017 IEEE/ACM 25th International Conference on Program Comprehension (ICPC)*, 2017, pp. 207–217.

[13] B. Lin, G. Robles, and A. Serebrenik, “Developer Turnover in Global, Industrial Open Source Projects: Insights from Applying Survival Analysis,” in *Proceedings of the 12th International Conference on Global Software Engineering*, 2017, pp. 66–75.

[14] M. Zhou, A. Mockus, X. Ma, L. Zhang, and H. Mei, “Inflow and Retention in OSS Communities with Commercial Involvement,” *ACM Trans. Softw. Eng. Methodol.*, vol. 25, no. 2, pp. 1–29, Apr. 2016.

[15] A. Schilling, “What Do We Know about FLOSS Developers’ Attraction, Retention, and Commitment? A Literature Review,” in *2014 47th Hawaii International Conference on System Sciences*, 2014, pp. 4003–4012.

[16] V. Midha and P. Palvia, “Retention and Quality in Open Source Software Projects,” in *Americas Conference on Information Systems*, 2007.

[17] A. Schilling, S. Laumer, and T. Weitzel, “Who Will Remain? An Evaluation of Actual Person-Job and Person-Team Fit to Predict Developer Retention in FLOSS Projects,” in *2012 45th Hawaii International Conference on System Sciences*, 2012, pp. 3446–3455.

[18] A. Barcomb and Ann, “Volunteer Attraction and Retention in Open Source Communities,” in *Proceedings of The International Symposium on Open Collaboration - OpenSym ’14*, 2014, pp. 1–2.

[19] K. Yamashita, Y. Kamei, S. Mcintosh, A. E. Hassan, and N. Ubayashi, “Magnet or Sticky? Measuring Project Characteristics from the Perspective of Developer Attraction and Retention,” *J. Inf. Process.*, vol. 24, pp. 339–348, 2016.

[20] M. Foucault, M. Palyart, X. Blanc, G. C. Murphy, and J.-R. Falleri, “Impact of developer turnover on quality in open-source software,” in *Proceedings of the 2015 10th Joint Meeting on Foundations of Software Engineering - ESEC/FSE 2015*, 2015, pp. 829–841.

[21] K. Nakakoji, Y. Yamamoto, Y. Nishinaka, K. Kishida, and Y. Ye, “Evolution patterns of open-source software systems and communities,” in *Proceedings of the international workshop on Principles of software evolution - IWPSE ’02*, 2002, p. 76.

[22] I. Steinmacher, M. A. Gerosa, and D. F. Redmiles, “Attracting, Onboarding, and Retaining Newcomer Developers in Open Source Software Projects,” in *Workshop on Global Software Development in a CSCW Perspective held in conjunction with the17th ACM Conference on Computer Supported Cooperative Work & Social Computing (CSCW{\textquoteright}14)*, 2014, no. February, pp. 1–4.

[23] C. Hannebauer and V. Gruhn, “On the Relationship between Newcomer Motivations and Contribution Barriers in Open Source Projects,” in *Proceedings of the 13th International Symposium on Open Collaboration - OpenSym ’17*, 2017, pp. 1–10.

[24] G. Von Krogh, S. Haefliger, S. Spaeth, and M. W. Wallin, “Carrots and rainbows: motivation and social practice in open source software development,” *MIS Q.*, vol. 36, no. 2, pp. 649–676, 2012.

[25] C. Santos, G. Kuk, F. Kon, and J. Pearson, “The attraction of contributors in free and open source software projects,” *J. Strateg. Inf. Syst.*, vol. 22, no. 1, pp. 26–45, Mar. 2013.

[26] I. Steinmacher, T. Conte, M. A. Gerosa, and D. Redmiles, “Social Barriers Faced by Newcomers Placing Their First Contribution in Open Source Software Projects,” in *Proceedings of the 18th ACM Conference on Computer Supported Cooperative Work & Social Computing - CSCW ’15*, 2015, pp. 1379–1392.

[27] L. Williams and R. R. Kessler, "Pair programming illuminated." Addison-Wesley, 2003.

[28] S. Chengalur-Smith and A. Sidorova, “Survival of Open-Source Projects: A Population Ecology Perspective,” 2003.

[29] I. Samoladas, L. Angelis, and I. Stamelos, “Survival analysis on the duration of open source projects,” *Inf. Softw. Technol.*, vol. 52, no. 9, pp. 902–922, Sep. 2010.

[30] J. Wang, “Survival factors for Free Open Source Software projects: A multi-stage perspective,” *Eur. Manag. J.*, vol. 30, no. 4, pp. 352–371, Aug. 2012.

[31] J. F. Low, T. Yathog, and D. Svetinovic, “Software analytics study of Open-Source system survivability through social contagion,” in *2015 IEEE International Conference on Industrial Engineering and Engineering Management (IEEM)*, 2015, pp. 1213–1217.

[32] E. Constantinou and T. Mens, “An empirical comparison of developer retention in the RubyGems and npm software ecosystems,” *Innov. Syst. Softw. Eng.*, vol. 13, no. 2–3, pp. 101–115, Sep. 2017.

[33] D. Ehls, “Open Source Project Collapse-Sources and Patterns of Failure,” in *Proceedings of the 50th Hawaii International Conference on System Sciences*, 2017, pp. 5327–5336.

[34] J. Coelho and M. T. Valente, “Why modern open source projects fail,” in *Proceedings of the 2017 11th Joint Meeting on Foundations of Software Engineering - ESEC/FSE 2017*, 2017, pp. 186–196.

[35] S. K. Shah, “Motivation, Governance, and the Viability of Hybrid Forms in Open Source Software Development,” *Manage. Sci.*, vol. 52, no. 7, pp. 1000–1014, Jul. 2006.

[36] H. Zhu, A. Zhang, J. He, R. E. Kraut, and A. Kittur, “Effects of peer feedback on contribution,” in *Proceedings of the SIGCHI Conference on Human Factors in Computing Systems - CHI ’13*, 2013, p. 2253.